\documentclass[manuscript]{aastex62}

\usepackage{amsmath} 
\received{2019 August 12}
\revised{2019 August 30}
\accepted{2019 September 18}
\submitjournal{ApJL}

\shorttitle{Sample article}
\shortauthors{Wu et al.}
\begin{document}

\title{Dependence of 3-D Self-correlation Level Contours on the Scales in the Inertial Range of Solar Wind Turbulence}

\correspondingauthor{Chuanyi Tu}
\email{chuanyitu@pku.edu.cn}

\author{Honghong Wu}
\affiliation{School of Earth and Space Sciences, Peking University, Beijing, China}

\author{Chuanyi Tu}
\affiliation{School of Earth and Space Sciences, Peking University, Beijing, China} 

\author{Xin Wang}
\affiliation{School of Space and Environment, Beihang University, Beijing, China} 

\author{Jiansen He}
\affiliation{School of Earth and Space Sciences, Peking University, Beijing, China}

 \author{ Linghua Wang}
\affiliation{School of Earth and Space Sciences, Peking University, Beijing, China} 
 


\begin{abstract}
The self-correlation level contours at $10^{10}\  \mathrm{cm}$ scale reveal a 3-D isotropic feature in the slow solar wind and a quasi-anisotropic feature in the fast solar wind. However, the $10^{10}\  \mathrm{cm}$ scale is approximately near the low-frequency break (outer scale of turbulence cascade), especially in the fast wind. How the self-correlation level contours behave with dependence on the scales in the inertial range of solar wind turbulence remains unknown. Here we present the 3-D self-correlation function level contours and their dependence on the scales in the inertial range for the first time. We use data at 1 AU from instruments on Wind spacecraft in the period 2005-2018. We show the 3-D isotropic self-correlation level contours of the magnetic field in the inertial range of both slow and fast solar wind turbulence. We also find that the  self-correlation level contours of the velocity in the inertial range present 2-D anisotropy with an elongation in the perpendicular direction and 2-D isotropy in the plane perpendicular to the mean magnetic field. These results present differences between the magnetic field and the velocity, providing new clues to interpret the solar wind turbulence in the inertial scale. 
%The self-correlation level contours at hour timescales reveal a 2-D isotropic feature in both the slow solar wind fluctuations and the fast solar wind fluctuations. However, this  isotropic feature is based on the assumption of axisymmetry with respect to the mean magnetic field. Whether the 3-D feature of the solar wind turbulence remains isotropic is still unknown. Here we perform the first three-dimensional self-correlation level contours analysis on the solar wind turbulence feature. We construct a 3-D coordinate system based on the maximum fluctuation direction identified by the minimum-variance analysis (MVA) method and the mean magnetic field direction.
%We apply the minimum-variance analysis (MVA) method to find the maximum fluctuation direction and combine it with the mean magnetic field direction to construct a 3-D coordinate system. We use intervals with 1 hour duration observed by WIND spacecraft from 2005 to 2018. We find that in the slow solar wind, the self-correlation level contour surfaces for both magnetic field and velocity field are almost spherical, which indicates a 3-D isotropic feature. We also find less spherical contour surfaces in the fast wind with a weak elongation in one of the perpendicular directions $r_{\mathrm{\perp 2}$. The 3-D feature of contour surfaces in the solar wind turbulence cannot be explained by the existed theory.

\end{abstract}

\keywords{Solar wind, Interplanetary turbulence, Magnetic fields, Space plasmas}


\section{Introduction} \label{sec:intro}

The magnetic field and the velocity both display broad-band fluctuations in the solar wind. The ubiquitous observation of Kolmogorov-like magnetic and velocity power spectra suggests the existence of the turbulent energy cascade in the inertial range \citep{Frisch1995, Tu1995SSRv, Bruno2013LRSP}. At the large scale side of the inertial range, the fast solar wind often presents a robust $1/f$ scaling injection range \citep{ Horbury1996AA, Matthaeus2007ApJ, Bruno2013LRSP} for both the magnetic field and the velocity. The low-frequency break between the inertial range and the injection range is around $10^{-3}$ Hz, a typical value for the fast wind at $1AU$ \citep{Bruno2013LRSP, Bruno2019aa}. In the slow solar wind, the $1/f$ scaling is also present for the magnetic field with a smaller low-frequency break around $10^{-4}$ Hz. However, the velocity spectrum keeps the Kolmogorov-like scaling throughout the analyzed frequency range by \cite{Bruno2019aa}.

\cite{Matthaeus1990JGR} developed the 2-D self-correlation function method to study the solar wind turbulence and obtained the famous “Maltese cross”. \cite{Dasso2005ApJ} applied the same method and found that the anisotropy behaves differently for the slow wind and the fast wind shown by the self-correlation function level contours analyzing the two-day intervals measured by Advanced Composition Explorer (ACE) spacecraft at $1AU$. 2D self-correlation function is also constructed by analyzing the simultaneous measurements from Cluster 4-spacecraft, showing anisotropic characteristics at small scales close to ion kinetic scales for both solar wind and magnetosheath turbulences \citep{Osman2006ApJ, He2011JGR}). \cite{Wang2019aApJ} extended this study using the same data set and found that the anisotropy disappears and it becomes 2-D isotropic for both the slow wind and the fast wind with the intervals duration decreasing from 2 days, 1 day, 10 hours, 2 hours, 1 hour. \cite{Wu2019ApJ} further extended the self-correlation function level contours analysis using 1-hour intervals observed by WIND spacecraft. They show a 3-D isotropic self-correlation function level contours in the slow wind and a 3-D quasi-isotropic self-correlation function level contours in the fast wind. The contour scale of 1-hour intervals is around $10^{10}\  \mathrm{cm}$ nearby the low-frequency break scale.  However, the feature of the self-correlation function level contours and its dependence on the scales in the inertial range remains unknown.  
  
In the present study, we perform the 3-D self-correlation function level contour analysis on the WIND spacecraft measurements using intervals with durations = 1 hour, 30 minutes and 10 minutes. We briefly introduce the method in section \ref{sec:DATA} and present our observational results in section \ref{sec:RESULTS}.  In section \ref{sec:CONCLUSIONS}, we draw our conclusions.
 
\section{DATA AND METHOD} \label{sec:DATA}
We briefly describe the method used in the analysis, more details can be found in \cite{Wu2019ApJ}. We use the magnetic field data with a cadence of $\Delta =3$ s from the magnetic field investigation \citep{Lepping1995SSRv} and the plasma data with a same time resolution $3 s$ from the three-dimensional plasma analyzer \citep{Lin1995SSRv} on board the WIND spacecraft in the period $2005-2018$. The data set was cut into intervals with duration of $T$, where $T=$ 1 hour, 30 minutes, and 10 minutes, respectively. These intervals were conserved for further investigation which contain less than $5\%$ data gap and $max[|\delta B_j|]<2  \mathrm{nT}$, $ \ max[|\delta V_j|]<20 \mathrm{km}$, where $j$ indicates $x$, $y$, $z$ component in the geocentric-solar-ecliptic (GSE) coordinate system, and $\delta$ means the variation between every $3$ s. 

The two-time-point self-correlation function for each interval $i$ is defined as
\begin{equation}
R_{U}(i,\tau)=<\delta  \vec{U}(t)\cdot\delta \vec{U}(t+\tau)>,
\end{equation}    
where $\tau=0, \Delta,2\Delta,...,T/2$ is the time lag, $<>$ denotes an ensamble time average, $\delta  \vec{U}$ is the time series removing a linear trend for either magnetic field $ \vec{B}$ or velocity $ \vec{V}$. We normalize the self-correlation function using the zero time lag self-correlation $R_{uu}(i,\tau)=R_{U}(i,\tau)/R(i, 0)$. We further obtain the spatial lag $r$ using Taylor hypothesis \citep{Taylor1938RS} $r=\tau V_{\mathrm{SW}}$, where $V_{\mathrm{SW}}$ is the mean flow velocity in the corresponding interval $i$. 

The 3-D coordinate system is constructed using the mean magnetic field $\vec{B}_\mathrm{0}$ and the maximum variance direction $L$ obtained by minimum-variance analysis (MVA) method \citep{Sonnerup1967JGR}. The $r_{\mathrm{\parallel}}$ and $r_{\mathrm{\perp 2}}$ components are defined as the mean magnetic field $\vec{B}_\mathrm{0}$ and the projection of $L$ in the plane perpendicular to $\vec{B}_\mathrm{0}$. $r_{\mathrm{\perp 1}}=r_{\mathrm{\parallel}}\times r_{\mathrm{\perp 2}}$. We calculate the angle $\theta_{\mathrm{VB}}$ between $\vec{V_{\mathrm{SW}}}$ and $\vec{B}_\mathrm{0}$ and the angle $\phi_{\mathrm{L}}$ between  $r_{\mathrm{\perp 2}}$ direction and the component of $\vec{V_{\mathrm{SW}}}$ perpendicular to $\vec{B}_\mathrm{0}$ for each interval $i$. 

We divide these intervals into the slow wind ($V_{\mathrm{SW}}<400$ km/s ) and the fast wind ($V_{\mathrm{SW}}>500$ km/s ) and study their 3-D self-correlation level contours separately. For $T=$ 30 minutes, we obtain $55331$ intervals in the slow wind and $10733$ intervals in the fast wind. For $T=$ 10 minutes, the numbers are $217830$ and $63656$. $\theta_{\mathrm{VB}}$ and $\phi_{\mathrm{L}}$ are binned into $15^\circ$ bins and the average of the normalized self-correlation functions is calculated as 
\begin{equation}
R_{uu}(\theta_{\mathrm{VB}}^m,\phi_{\mathrm{L}}^n,r)=\frac{1}{n(\theta_{\mathrm{VB}}^m,\phi_{\mathrm{L}}^n)} \sum_{\substack{\theta_{\mathrm{VB}}^m-7.5<=\theta_{\mathrm{VB}}(i)<\theta_{\mathrm{VB}}^m+7.5,\\ \phi_{\mathrm{L}}^n-7.5<=\phi_{\mathrm{L}}(i)<\phi_{\mathrm{L}}^n+7.5}}R_{uu}(i,r),
\end{equation} 
where $n(\theta_{\mathrm{VB}}^m,\phi_{\mathrm{L}}^n)$ denotes the number of the intervals in corresponding bin, and, $\theta_{\mathrm{VB}}^m=15^\circ m+7.5^\circ;\phi_{\mathrm{L}}^n=15^\circ n+7.5^\circ; m,n=0,1,2,...,5.$
  
\begin{figure}[ht!]
\includegraphics[width=\linewidth]{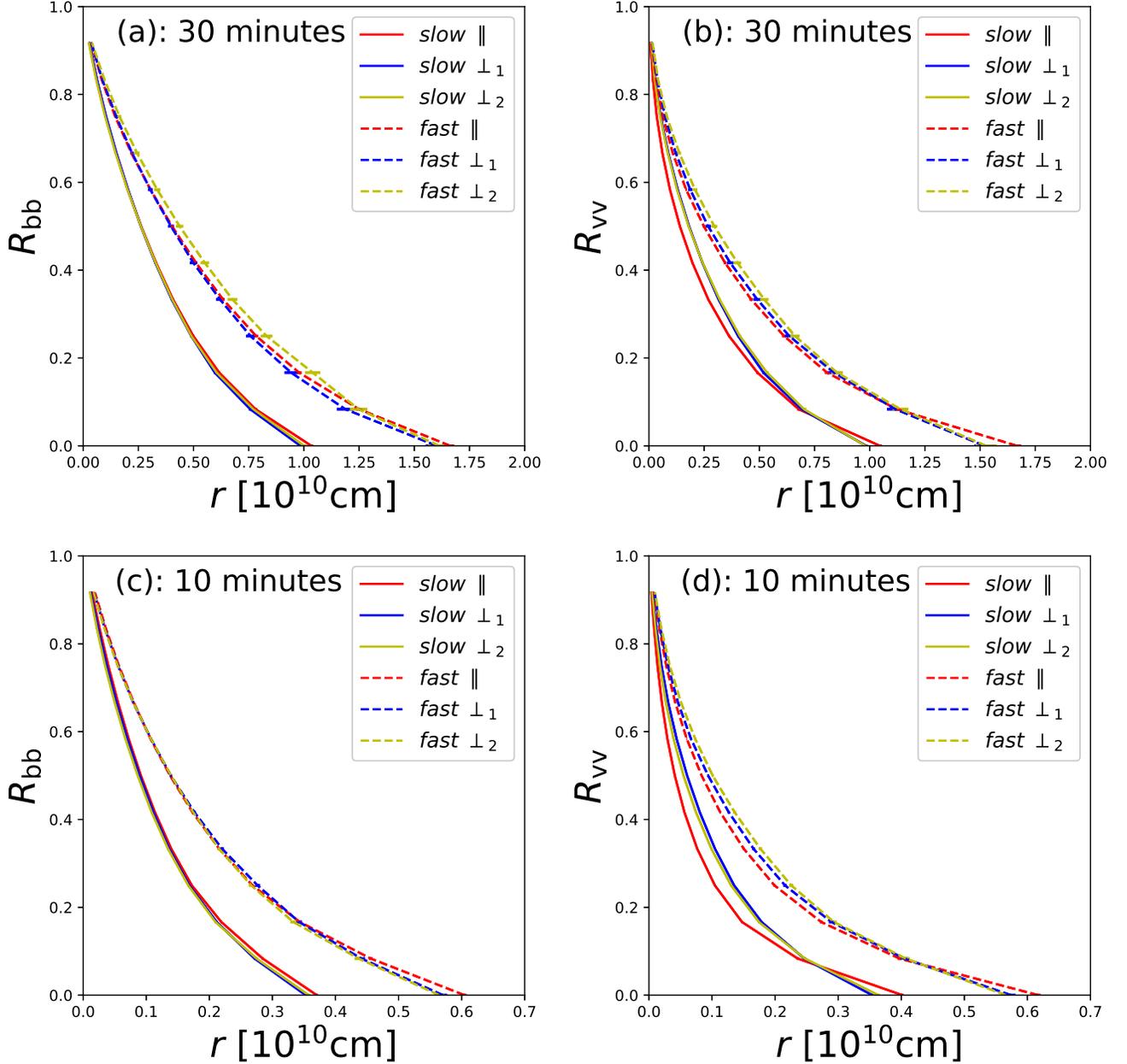}
\caption{(a): Averaged normalized self-correlation functions $R_{\mathrm{bb}}(r) $ of 30-minute magnetic field data with the standard error bars of $r_{\mathrm{level}}$ for a given $R_{\mathrm{bb}}$. The solid (dashed) lines are for the slow (fast) wind. Red, blue,  and yellow colors indicate the $r_{\mathrm{\parallel}}$, $r_{\mathrm{\perp 1}}$, and $r_{\mathrm{\perp 2}}$ directions, respectively. (b): Averaged normalized self-correlation functions $R_{\mathrm{vv}}(r) $ of 30-minute-long velocity data, in the same manner as in (a). (c): Same as in (a) for 10-minute magnetic field data. (d):  Same as in (b) for 10-minute velocity data.}\label{fig:figure1}
\end{figure}

We obtain 36 averaged self-correlation functions for 36 $(\theta_{\mathrm{VB}},\phi_{\mathrm{L}})=15^\circ \times 15^\circ$ bins. Figure \ref{fig:figure1} shows the averaged self-correlation functions for $T=$  30 minutes and $T=$  10 minutes in the $r_{\mathrm{\perp 1}}\ (75^\circ<=\theta_{\mathrm{VB}}<=90^\circ,\ 75^\circ<=\phi_{\mathrm{L}}<=90^\circ)$, $r_{\mathrm{\perp 2}}\ (75^\circ<=\theta_{\mathrm{VB}}<=90^\circ,\ 0^\circ<=\phi_{\mathrm{L}}<15^\circ)$, and $r_{\mathrm{\parallel}}\ (0^\circ<=\theta_{\mathrm{VB}}<15^\circ,\ 0^\circ<=\phi_{\mathrm{L}}<=90^\circ)$ directions.

In the left panels of Figure \ref{fig:figure1}, we present the averaged magnetic self-correlation functions with standard error bars in both the slow wind (solid lines) and the fast wind (dashed lines). The functions of the three directions are almost overlapped with each other for both 30 minutes and 10 minutes intervals, especially for the slow wind, indicating the isotropy of the self-correlation functions. In the right panels of Figure \ref{fig:figure1}, we show the averaged self-correlation functions of the velocity with standard error bars for both the slow wind and the fast wind. The velocity functions of the three directions have more difference with each other than the magnetic field. In general, the parallel function is smaller than both of the perpendicular functions and the perpendicular functions behave similar with each other. For both the magnetic field and the velocity, the correlation functions of the slow wind decrease more rapidly than that of the fast wind for both 30 minutes and 10 minutes intervals. This difference between the slow and fast wind has already been shown for 1 hour intervals in \cite{Wu2019ApJ}, where $23083$ intervals in the slow wind and $3347$ intervals in the fast wind are investigated.

For each bin $(\theta_{\mathrm{VB}},\phi_{\mathrm{L}})$, we calculate $r$ at level  $R_{uu}(\theta_{\mathrm{VB}},\phi_{\mathrm{L}},r)=1/e \approx 0.368$ and denote the result as $r_{\mathrm{level}}$. We transform $(\theta_{\mathrm{VB}},\phi_{\mathrm{L}},r_{\mathrm{level}})$ into $(r_{\mathrm{\perp 1},} r_{\mathrm{\perp 2}}, r_{\mathrm{\parallel}})$ as 
\begin{align}
r_{\perp1}&=r_{\mathrm{level}}\sin\theta_{\mathrm{VB}}\sin\phi_{\mathrm{L}},\\
r_{\mathrm{\perp 2}}&=r_{\mathrm{level}}\sin\theta_{\mathrm{VB}}\cos\phi_{\mathrm{L}},\\
r_{\mathrm{\parallel}}&=r_{\mathrm{level}}\cos\theta_{\mathrm{VB}}.
\end{align}  
We also obtain $r_{\mathrm{level}}(i)$ for each interval $i$ at level  $R_{uu}(i,r)=1/e \approx 0.368$. We define two ratios $r_{\mathrm{\parallel}}^c/r_{\mathrm{\perp}}^c$ and $r_{\mathrm{\perp2}}^c/r_{\mathrm{\perp1}}^c$, where 
\begin{equation}
r_{\mathrm{\parallel}}^c =\frac{1}{n(r_{\mathrm{\parallel}}^c)}\sum_{\substack{0<=\theta_{\mathrm{VB}}(i)<15,\\ 0<=\phi_{\mathrm{L}}(i)<90}} r_{\mathrm{level}} (i),
\end{equation} 
\begin{equation}
r_{\mathrm{\perp}}^c =\frac{1}{n(r_{\mathrm{\perp}}^c )}\sum_{\substack{75<=\theta_{\mathrm{VB}}(i)<90,\\ 0<=\phi_{\mathrm{L}}(i)<90}} r_{\mathrm{level}} (i),
\end{equation} 
\begin{equation}
r_{\mathrm{\perp2}}^c =\frac{1}{n(r_{\mathrm{\perp2}}^c)}\sum_{\substack{60<=\theta_{\mathrm{VB}}(i)<90,\\ 0<=\phi_{\mathrm{L}}(i)<15}} r_{\mathrm{level}} (i),
\end{equation} 
and,
\begin{equation}
r_{\mathrm{\perp1}}^c =\frac{1}{n(r_{\mathrm{\perp1}}^c )}\sum_{\substack{60<=\theta_{\mathrm{VB}}(i)<90,\\ 75<=\phi_{\mathrm{L}}(i)<90}} r_{\mathrm{level}} (i).
\end{equation}  
The ratio $r_{\mathrm{\parallel}}^c/r_{\mathrm{\perp}}^c$ describes the $r_{\mathrm{level}}$ difference between the parallel and the perpendicular direction, and the ratio $r_{\mathrm{\perp2}}^c/r_{\mathrm{\perp1}}^c$ describes the anisotropy in the perpendicular plane. The result is shown in the next section.

\section{RESULTS} \label{sec:RESULTS}

\begin{figure}[ht!]
\includegraphics[width=\linewidth]{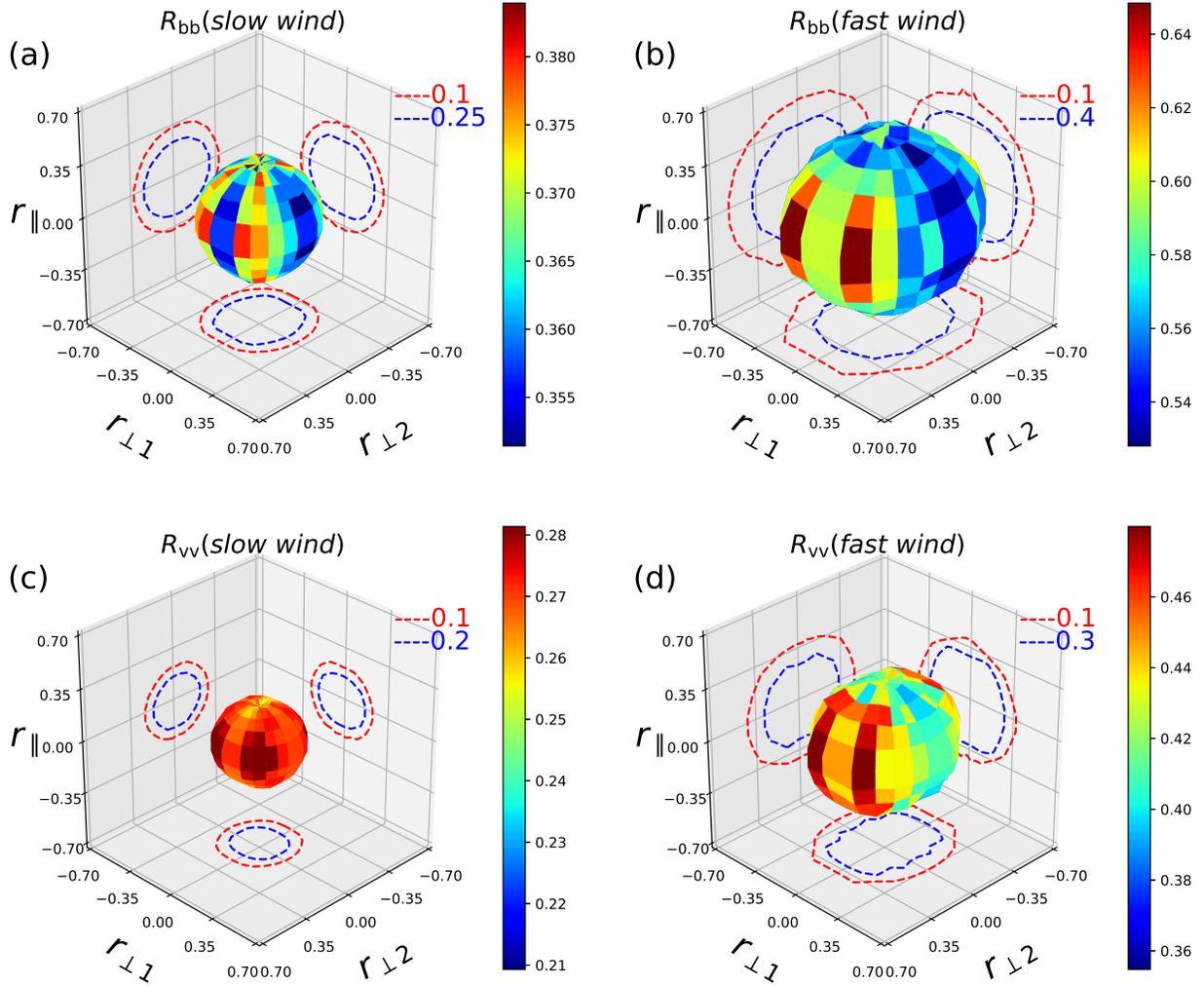}
\caption{ 3-D self-correlation level contour surface at level $R_{uu}=0.368$ of the 30-minute-long data for the (a) magnetic field in the slow wind; (b) magnetic in the fast wind; (c) velocity in the slow wind; (d) velocity field in the fast wind. The color represents the distances from the origin $r_{\mathrm{level}}\  [10^{10}\ \mathrm{cm}]$. The dashed red (blue) lines in $r_{\mathrm{\perp 1}}=-0.70$ plane are projections of the intersection lines of the surface with two planes $r_{\mathrm{\perp 1}}=A1\ (A2)$, where $A1$ and $A2$ are shown in the legends with the corresponding colors in the corresponding panel, same for the other two planes.}\label{fig:figure2}
\end{figure}

Figure \ref{fig:figure2} shows 3-D self-correlation level contour surfaces at level $R_{uu}=0.368$ for 30-minute intervals. The spatial lag scale for $R_{uu}=0.368$ is around $ 10^{9}\ \mathrm{cm}$ for 30 minutes, which is in the inertial scale of solar wind turbulence at $1$ AU. In Figure \ref{fig:figure2}(a), the slow wind magnetic field self-correlation level contour surface is spherical. The projection closed curves on the 2-D plane are plotted to help visualize the 3-D feature. In Figure \ref{fig:figure2}(b), the fast wind magnetic field self-correlation level contour surface shows a weak elongation along $r_{\mathrm{\perp 2}}$. In Figure \ref{fig:figure2}(c), the slow wind velocity self-correlation level contour surface presents a weak elongation in the perpendicular plane. In Figure \ref{fig:figure2}(d), the fast wind velocity field self-correlation level contour surface has a similar shape with that of magnetic field. The size difference of the surface remains for 30 minutes intervals as for 1 hour intervals shown in \cite{Wu2019ApJ}: the $r_{\mathrm{level}}$ is longer for the fast wind than for the slow wind and longer for the magnetic field than for the velocity. We also analyzed the 3-D self-correlation level contours in the LMN coordinate system, constructed by MVA analysis: L is the directions of the maximum variance as the one used for the construction of the 3-D coordinate system described in Section \ref{sec:DATA}, N is the directions of the minimum variance, and M is directions of the immediate variance. The 3-D features of self-correlation level contours in the LMN coordinate system (not shown) are the same as we shown in Figure \ref{fig:figure2}.
 
%\begin{figure}[ht!]
%\includegraphics[width=\linewidth]{dependence_figure2.eps}
%\caption{ 3-D self-correlation level contour surface at level $R_{uu}=0.368$ of the 10-minute-long data for (a) magnetic field in the slow wind; (b) magnetic field in the fast wind; (c) velocity field in the slow wind; (d) velocity field in the fast wind. The color represents $r_{\mathrm{level}}\  [10^{10}\ \mathrm{cm}]$, which is the distances from the origin. The dashed red and blue lines in $r_{\mathrm{\perp 1}}=-0.70$ plane are projections of the intersection lines of the surface with two planes $r_{\mathrm{\perp 1}}=A1$ and  $r_{\mathrm{\perp 1}}=A2$, respectively, where $A1$ and $A2$ are shown in the legends with the corresponding colors in the corresponding panel ; the dashed red and blue lines in $r_{\mathrm{\perp 2}}=-0.70$ plane are projections of the intersection lines of the surface with two plane $r_{\mathrm{\perp 2}}=A1$ and  $r_{\mathrm{\perp 2}}=A2$, respectively; the dashed red and blue lines in $r_{\mathrm{\parallel}}=-0.70$ plane are projections of the intersection lines of the surface with two plane $r_{\mathrm{\parallel}}=A1$ and  $r_{\mathrm{\parallel}}=A2$, respectively. }\label{fig:figure2}
%\end{figure}

   \begin{figure}[ht!]
\includegraphics[width=\linewidth]{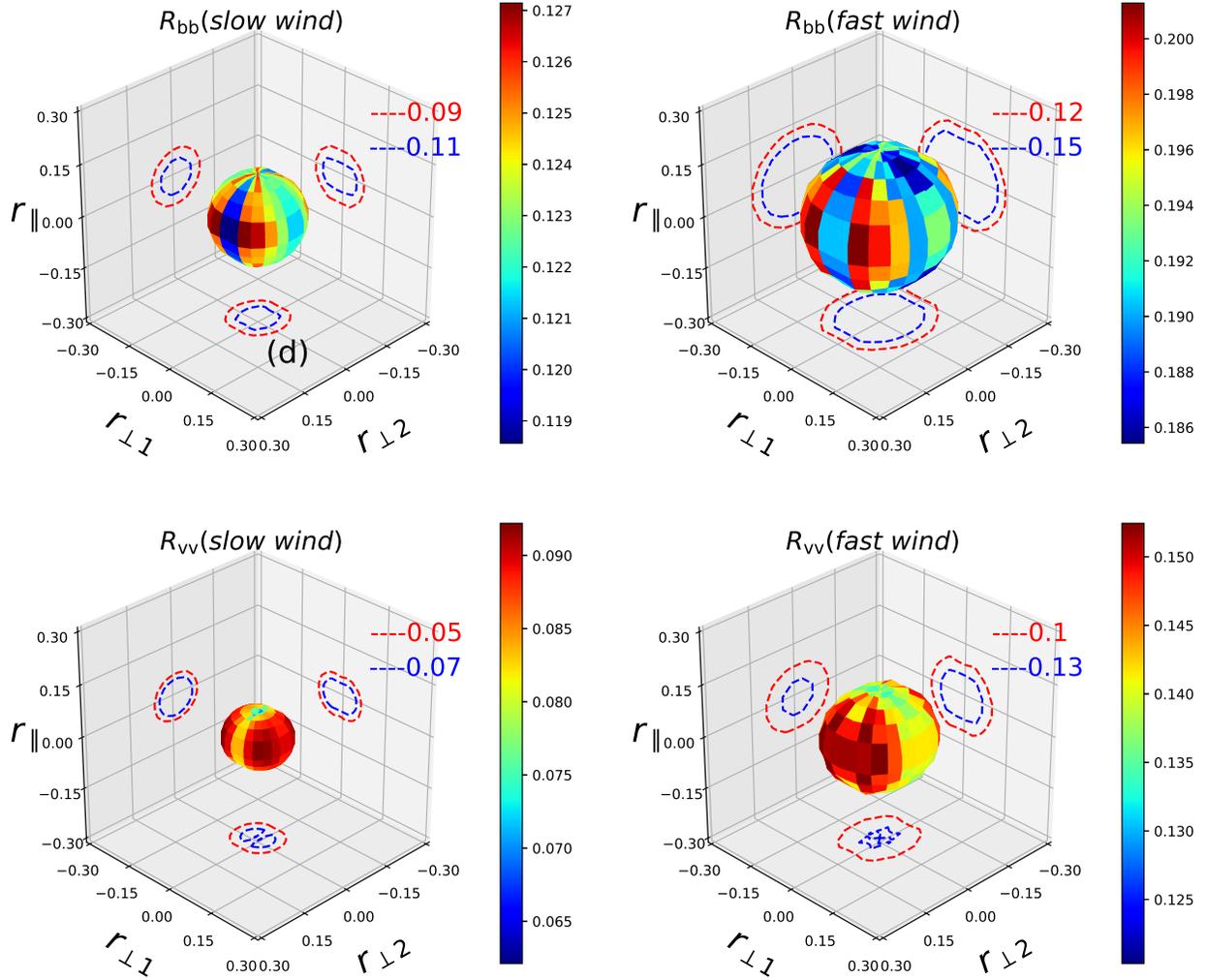}
\caption{ 3-D self-correlation level contour surface at level $R_{uu}=0.368$ of the 10-minute data with the same panel and line styles as Figure \ref{fig:figure2}, except that the projection planes are  $r_{\mathrm{\parallel}}=-0.30$, $r_{\mathrm{\perp 1}}=-0.30$ and $r_{\mathrm{\perp 2}}=-0.30$.}\label{fig:figure3}
\end{figure}
Figure \ref{fig:figure3} shows 3-D self-correlation level contour surfaces at level $R_{uu}=0.368$ for 10-minute intervals. The size of surface reaches $ 6\times 10^{8}\ \mathrm{cm}$ in the slow wind for the velocity. The magnetic field self-correlation level contour surfaces for both the slow wind and the fast wind are almost spherical, suggesting the isotropy for the magnetic field as shown in Figure \ref{fig:figure3}(a) and \ref{fig:figure3}(b). The elongation in the perpendicular plane of the velocity field self-correlation function contour surfaces grows in both the slow wind and the fast wind seen in Figure \ref{fig:figure3}(c) and \ref{fig:figure3}(d).

\begin{figure}[ht!]
\includegraphics[width=\linewidth]{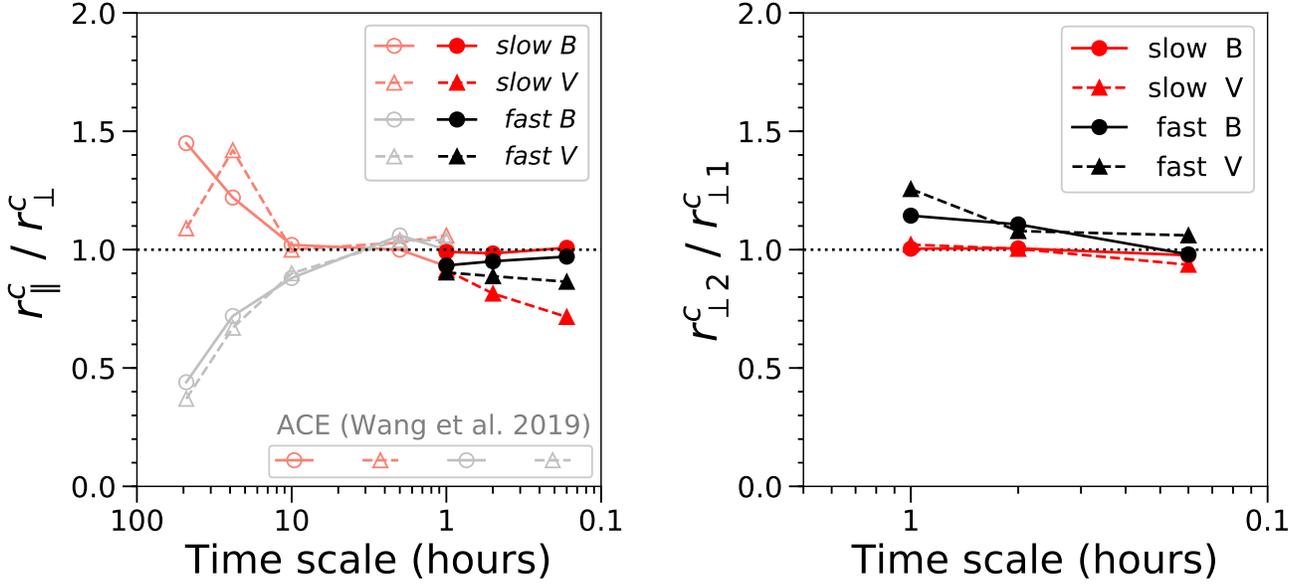}
\caption{Left panel: $r_{\mathrm{\parallel}}^c/r_{\mathrm{\perp}} ^c$ for intervals with different duration. The red (black) lines are for the slow (fast) wind. The solid (dashed) lines indicate the magnetic field (velocity) results. From left to right, the timescales correspond to 2 days, 1 day, 10 hours, 2 hours, 1 hour, 30 minutes, and 10 minutes, respectively. The ratios shown in hollow circles and hollow triangles are from \cite{Wang2019aApJ} with ACE measurements at level=$0.8$. While the solid circles and solid triangles are from our WIND observations at level $R_{uu}=0.368$. Right panel: $r_{\mathrm{\perp2}}^c/r_{\mathrm{\perp1}}^c$ for intervals with different duration at level $R_{uu}=0.368$ with the same line styles as in the left panel. From left to right, the timescales correspond to 1 hour, 30 minutes, and 10 minutes, respectively. }\label{fig:figure4}
\end{figure}

In the left panel of Figure \ref{fig:figure4}, we show $r_{\mathrm{\parallel}}^c/r_{\mathrm{\perp}}^c$ with interval durations = 1 hour, 30 minutes and 10 minutes for the magnetic field (solid circles) and the velocity field (solid triangles) measured by Wind spacecraft for both the slow (red) and fast (black) winds. For comparison, the results of ACE observations from \cite{Wang2019aApJ} at level $R_{uu}=0.8$ with time durations = 2 days, 1 day, 10 hours, 2 hours, 1 hour are shown here in hollow circles and hollow triangles. The results of 1 hour intervals in our work are not exactly the same as in \cite{Wang2019aApJ}. That may attribute to the different data sets. The $r_{\mathrm{\parallel}}^c/r_{\mathrm{\perp}}^c$ of the magnetic field in both the slow wind and the fast wind with interval durations = 1 hour, 30 minutes and 10 minutes are all very close to $1$, indicate a 2-D isotropy shown by the self-correlation level contour in the inertial scale. The ratios of the velocity field are less than $1$, especially in the slow wind. They are $0.72$ and $0.86$ for 10 minutes intervals in the slow wind and fast wind obtained by averaging $r_{\mathrm{level}}$. We calculate the ratios directly using the $r_{\mathrm{level}}$ calculated by the correlation functions shown in Figure \ref{fig:figure1} at $R_{vv}=0.368$ and they are $0.73$ and $0.86$. These results suggest an elongation along the perpendicular direction for the velocity in the inertial scale. The right panel of Figure \ref{fig:figure4} shows  $r_{\mathrm{\perp2}}^c/r_{\mathrm{\perp1}}^c$, suggesting the isotropy in the perpendicular plane for both the magnetic field and the velocity field in the slow wind. In the fast wind, the elongations along $r_{\mathrm{\perp2}}$ for 1 hour disappear for 10 minutes, indicating the isotropy in the perpendicular plane deeply into the inertial scale.
%The anisotropy shown by the contours at different levels are scale-independent \citep{Wang2019aApJ} and therefore we can compare the ratio at level $R_{uu}=0.8$ and $R_{uu}=0.368$. 
 
 \section{discussion and CONCLUSIONS} \label{sec:CONCLUSIONS}
 
 We present the 3-D self-correlation level contours and their dependence on the time scale in the inertial range using WIND measurements at 1 AU during 14 years from 2005 to 2018. We analyze the self-correlation level contours of the magnetic field and the velocity in both the slow wind and the fast wind for intervals with durations $=$ 30 minutes and 10 minutes. We use two ratios $r_{\mathrm{\parallel}}^c/r_{\mathrm{\perp}}^c$ and $r_{\mathrm{\perp2}} ^c/r_{\mathrm{\perp1}} ^c$ to describe the anisotropy and show their dependence on the time scale of the interval duration. The 3-D self-correlation level contours of the magnetic field present an isotropy in both the slow solar wind and the fast solar wind for both the 30-minute intervals and the 10-minute intervals, which corresponds to approximately scale $ 10^{9}\ \mathrm{cm}$ in the inertial scale. However, the 3-D self-correlation level contours of the velocity indicate an elongation in the direction perpendicular to the mean magnetic field, and 2-D isotropy in the plane perpendicular to the mean magnetic field. The behaviors of the magnetic field and the velocity and their differences are new. 
  
\cite{Carbone1995JGR} developed a model for the 3-D magnetic field correlation spectra and reconstructed the shape of the self-correlation level contours of "Maltese cross" using the minimum variance framework. Our new results are inconsistent with the "Maltese cross" and this inconsistency requires further study. It is also hard to understand our results under the framework of the critical balance theory \citep{Goldreich1995ApJ, Boldyrev2006PhRvL}, which predicts a strong anisotropy of the MHD turbulence spectrum consistent with some local observations of structure functions in the solar wind \citep{Chen2012ApJ, Verdini2018ApJ}. The existing theories cannot explain our results and we cannot provide an exhaustive explanation with our current level of understanding and numerical simulations. These results open a new window into interpreting the solar wind turbulence on the inertial scale. 
 
\acknowledgments

We thank the CDAWEB for access to the Wind data. This work at Peking University and Beihang University is supported by the National Natural Science Foundation of China under contract Nos. 41674171, 41874199, and 41574168, 41874200, 41774183, and 41861134033.

\end{document}